\documentclass[12pt,letterpaper]{article}
\usepackage{amsmath, amsthm, amsfonts, amssymb}
\usepackage{mathrsfs} 
\usepackage{hyperref}
\usepackage{graphicx}
\usepackage{braket}

\newcommand{\be}{\begin{equation}}
\newcommand{\ee}{\end{equation}}
\newcommand{\bea}{\begin{eqnarray}}
\newcommand{\eea}{\end{eqnarray}}
\def\ut{\tilde u}

\def\thintablerule{\hrule height0.4pt}

\allowdisplaybreaks

\begin{document}

\centerline
{\Huge Graph theoretical approach to conformal correlators:}
\centerline {\Huge The conformal squid}

\vskip 2 cm
\centerline{\large Nikos Irges and Stylianos Kastrinakis}
\vskip1ex
\vskip.5cm
\centerline{\it Department of Physics}
\centerline{\it National Technical University of Athens}
\centerline{\it Zografou Campus, GR-15780 Athens, Greece}
\vskip .4cm
\begin{center}
{\it e-mail: irges@mail.ntua.gr, skastrinakis@mail.ntua.gr}
\end{center}
\vskip 1.5 true cm
\centerline {\bf Abstract}
\vskip 3.0ex
\thintablerule
\vskip 2.0ex

We introduce a discrete, graph theoretical approach to conformal field theory correlators. In a certain basis, called the squid basis, the correlator
of $N$ scalar operators can be expressed as the determinant of a natural, conformally covariant metric on a weighted graph, called the squid graph.
We present the construction of this metric and discuss its possible role in constraining conformal data.

\vskip 2.0ex
\thintablerule

\vskip-0.2cm
\newpage

\section{Introduction}

\renewcommand*{\thefootnote}{\fnsymbol{footnote}}
\setcounter{footnote}{1}
\renewcommand*{\thefootnote}{\arabic{footnote}}
\setcounter{footnote}{0}

The term universality in quantum field theory encodes the observation that different physical systems, when they reach a ``fixed'' point under
their renormalization group flow where all $\beta$ functions vanish, are governed by the same conformal symmetry. 
It is then an important issue to understand to what extent universality is a physical or mathematical property.
The problem can be stated more concretely by recalling that the system's behavior near such fixed points is determined by the critical exponents
which are the quantum corrections to the classical scaling dimensions of operators $O_i$, forming the scaling dimensions $\Delta_i$.
These appear in the correlators
\be\label{Ncorrelator}
\Sigma_N(x_i;\Delta_i) = \langle O_1(x_1)\cdots O_N(x_N)\rangle
\ee
of operators.\footnote{For simplicity we will restrict the discussion to scalar operators.} These correlators are fixed in a conformal field theory (CFT) up to an invariant expression of conformal invariants; however, the conformal symmetry does not fix, at this level, the values of the scaling dimensions $\Delta_i$.

The question, therefore, is whether and by how far the underlying mathematical structure is able to constrain the $\Delta_i$ at a fixed point.
We know that in a strict CFT context \cite{CFT}, in order to constrain the scaling dimensions, at least one additional assumption is needed,
that of the operator product expansion (OPE), which is, however, a physical rather than a mathematical statement. 
Similarly, other physical constraints, such as the ones that originate from unitarity (reflection positivity) 
or geometric constraints that arise in one way or another after the OPE has been applied \cite{AHamed},
even though implicitly present, are not discussed here.

Instead, we would like to initiate an approach that could provide some new insight into the role that mathematics plays in the dynamics of conformal field theories. This will be done by gradually providing postulates about the mathematical framework being built through the use of a metric on a graph and looking at the consequences to the system's dynamics in the form of, ultimately, restrictions to its scaling dimensions $\Delta_i$. Building such a framework to be in line with everything known about conformal systems (such as the results of the OPE) while also looking for possible novel applications is, however, a vast endeavor, and as such we intent to perform it in a stepwise fashion. In this paper we present the derivation of the metric used and show how the $N$ point correlators derivable from it transform as expected under conformal symmetry. Afterward, we explain how one may draw parallels between the expression of the correlator in this form that uses the metric and the volume of a simplex that is associated with it. Finally, we also discuss possible future paths to follow.

\section{Conformal transformations on a graph}

We consider $N$ scalar operators $O_i$ of scaling dimension $\Delta_i$ placed in $D$-dimensional Euclidean space at points $x_i$.
The correlator between these operators depends only on the distances $x_{ij}\equiv |x_i-x_j|$ and the $\Delta_i$ for $N\le 3$
and for $N>3$ depends additionally on certain conformally invariant length ratios that we will denote as $u_{ij}$ in the following.
The approach we propose here is to view this as a discrete system that forms a graph with the operators replaced by the graph vertices and with 
the scaling dimensions playing the role of its vertex weights. The intervertex distances $x_{ij}$ are then the lengths of the edges of the graph
and can be considered as (geometric) edge weights.
The question we would like to ask is, what then is the quantity that replaces the correlator in this viewpoint?

\begin{figure}[h]
    \centering
    \includegraphics[width=\textwidth]{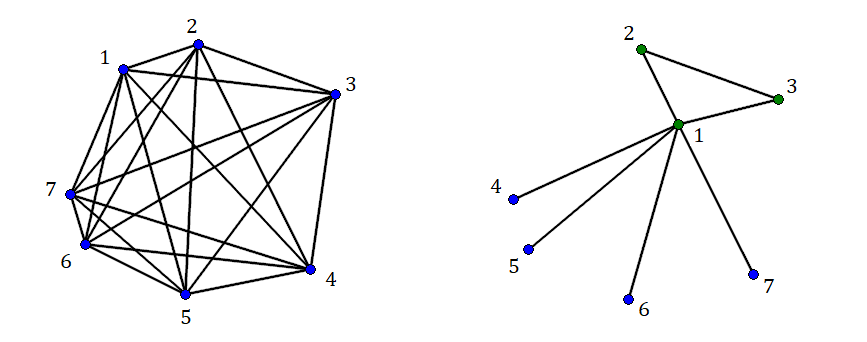}
    \caption{\label{fig:graphs}Left sketch: complete graph for seven points. Right sketch: squid graph, modulo index permutations, for seven points.}
\end{figure}

Given $N$ points, one can form various graphs with the points being the vertices of the graph. We will consider two types of simple graphs as 
being of particular interest to us, namely, the complete graph and the so-called squid graph; see \cite{bib:S2}. So, a squid graph $Sq(n,m)$ 
may be defined as the simple graph one obtains when starting off with a cyclic graph $C_n$ and then choosing one of those vertices and drawing 
another $m \geq 1$ vertices that we only connect with the chosen vertex through other $m$ edges. In the following we will deal solely with squid 
graphs where $n=3$, which we will simply refer to as squid graphs of $N$ points, where $N=m+3$. Examples of both types of graphs described can be seen in Fig. \ref{fig:graphs}.
Given a graph one can place weight functions on its vertices or its edges or both. One can show \cite{bib:S1} that in order to perform a 
conformal transformation upon the edge weights $w_{ij}$ of a discrete graph, these edge weights should transform according to the expression
\begin{equation}
w_{ij}' = w_{ij} ~ e^{u(v_i)} e^{u(v_j)} ,
\label{general_conf_transform_of_weights_Jacobson}
\end{equation}
where $u$ is a (weight) function of the vertex which we will call the conformal factor. Edge weights are considered to be dimensionless, even when
they are interpreted as the geometric edge weights $x_{ij}$. Thus, lengths, areas, and volumes are all dimensionless. When passing to physical applications, one can always 
properly restore dimensionalities.

Conformal transformations, by definition, pose stronger constraints on the system than those of the symmetry group of Euclidean 
space\footnote{As a convention, distances mentioned in the following will be implied to be Euclidean, not Minkowski, as one can in general 
apply a Wick rotation to move between descriptions.} and, as a consequence of this, distances no longer are invariant under all transformations.
Instead, one can prove \cite{bib:S5} that the conformal invariants of a set of $N$ points (in position space) and their in-between distances are formed 
by taking the \textit{length cross ratios}, i.e., certain ratios of products of lengths. In the following we will use a specific set of $N(N-3)/2$ independent 
invariants for the aforementioned $N$ points, this choice of invariants taken from \cite{bib:S4} as it is tailored to the needs of our description:
\begin{align}
\begin{split}
u_{2k} &\equiv \frac{x_{23} x_{1k}}{x_{13} x_{2k}}  , \quad 4 \leq k \leq N , \\
u_{3k} &\equiv \frac{x_{23} x_{1k}}{x_{12} x_{3k}}  , \quad 4 \leq k \leq N , \\
u_{ij} &\equiv \frac{x_{23} x_{1i} x_{1j}}{x_{ij} x_{12} x_{13}}  , \quad 4 \leq i < j \leq N .
\end{split}
\label{complete_set_of_generalised_conformal_invariants}
\end{align}
Dimensionality conditions may then further restrain the independent invariants of a system for larger $N$, the number of which will be dependent on the available dimensions $D$.
The above choice of invariants characterizes what we will call the \textit{squid basis} \cite{bib:S4}, which is a convenient basis where the correlator of $N$ scalar operators
can be simply expressed. Notice that the squid basis is such that its cross ratios contain one and only one edge that belongs to the complete graph but not to the squid.
These are the \textit{diagonals} of the squid, and their corresponding cross ratios parametrize its conformal deformations. 
The existence of conformal invariants implies that configurations of points which differ in their set of invariants are immediately ruled out of 
being able to conformally transform among themselves. Configurations that are related by conformal transformations
we consider to belong in the same \textit{conformal class}. An important result regarding these conformal 
classes is that one can prove \cite{bib:S1} that, for discrete graphs, there always exists a \textit{unique representative} of a conformal class. That is, given 
an arbitrary graph, one is always able to transform in a conformal way to some other specific graph of some property which uniquely defines it among all 
other graphs of the same conformal class. In the following we will refer to the positions of the points of such a unique representative as the \textit{reference configuration} of the associated conformal class.

In addition, we are given a concrete reference configuration in \cite{bib:S1}; namely, it is the configuration whose graph's weights $\bar{w}$ satisfy
\begin{equation}
\forall k \in \{1, \dots ,N\} : \prod_{e_l \sim v_k} \bar{w}_{ik} = 1 ,
\label{Jacobson_prop_of_unique_rep}
\end{equation}
where $e_l \sim v_k$ denotes that the edge $e_l$ that is associated with the weight $\bar{w}_{ik}$ has the vertex $v_k$ as an end point, and thus the product runs over all edges adjacent to a given vertex. And when this equality holds for all vertices of a weighted graph we have a graph positioned at a reference configuration of this property.

Thus far, as well as in the following, we have denoted general weights with $w,w', \dots$, while we will use the notation $\bar{w}$ for the reference configuration of  \cite{bib:S1} applied to a complete graph and the notation $\tilde{w}$ for the same reference configuration on a squid graph, which we will call the \textit{unit squid configuration}. As we are interested only in squid graphs of the form $Sq(3,N-3)$, it is easy to see that this means that all weights for its reference configuration are unit sized.

In regard to the conformal transformation \eqref{general_conf_transform_of_weights_Jacobson}, note how any set of real values for the conformal 
factors defines a valid conformal transformation that transforms the edge weights. We are thus free to rescale these, each by a different in general
(in the following positive) number, say, $\Delta_i > 0$. These rescalings simply define an automorphism between conformal transformations,
\begin{equation}
u_i ~\leftrightarrow~ u_i' = \Delta_i u_i ,
\label{conformal_factor_rescaling}
\end{equation}
with $u_i \equiv u(v_i)$, i.e., the conformal factors introduced in \eqref{general_conf_transform_of_weights_Jacobson}.
The set where $u_i=0 ~ \forall ~ i$ is the identity element of the group of conformal transformations. What remains is to choose a conformal transformation 
as the scale of these transformations, i.e., where $u_i=1 ~ \forall ~ i$. Since the problem described is not really changed, we are free to choose arbitrarily. 
We choose this unit transformation to be the one defined by the underlying geometric length space. The benefit of this is that it allows us to study the conformal transformations of lengths as a limit of our system.

Looking at the subcase of the underlying conformal space for which we define the weights to be lengths, one can see that lengths transform conformally according to
\begin{equation}
x'_{ij} = \left| \frac{\partial x_i'}{\partial x_i} \right|^{\frac{1}{2D}} \left| \frac{\partial x_j'}{\partial x_j} \right|^{\frac{1}{2D}} x_{ij}
\label{conf_transform_of_lengths}
\end{equation}
although we are in the following going to adopt a more agnostic stance, one dependent only on this conformal transformation law upon a graph, and disregard
the continuous transformations from which we derived these. This means that now we will treat the various Jacobian determinants as a set of independent values 
chosen by us / given to us that encode the conformal transformation to be performed. For this reason, as well as for notational brevity, we define the notation
\be
j_i \equiv \left| \frac{\partial x_i'}{\partial x_i} \right|^{\frac{1}{2D}} ~,
\ee
which we will solely use in the following. Note that conformal transformations are orientation preserving and thus for all the vertices $j_i \in (0,+\infty)$ should hold.

However, this is not the most general conformal weight function on the edges, as it is described by a particular conformal transformation. Put another way: by 
knowing any two of the three sets of initial weights, final weights, and conformal factors, one can solve for the third. But this excludes all other conformal transformations 
which one may obtain by rescaling as before: it restrains the various physical systems we may wish to model. To reestablish this freedom, we can allow for an 
extra set of vertex weights. So, we will look at the case where we are also given a set of independent values, i.e., in CFT these are the scaling dimensions $\Delta_i$. By rescaling the conformal factors,
\begin{equation}
u_i = \ln{ j_i } ~\rightarrow~ u_i = \Delta_i \ln{ j_i }
\label{conformal_factor_rescaling_explicit}
\end{equation}
the conformal transformation of weights takes the form
\begin{equation}
w_{ij}' = J_i J_j w_{ij} ,
\label{conf_transform_of_weights_Jacobson}
\end{equation}
where now we have additionally defined
\be
J_i \equiv j_i^{\Delta_i} ,
\label{bigJdef}
\ee
the appropriate Jacobian factors defining the conformal transformations of the weighted graph. 
Thus, in the following, we will use the transformation rule of a general edge weight $w_{ij}$ determined by the $J_i$ 
together with the same transformation determined by the corresponding $j_i$ applied to the underlying geometric length space.

\section{A determinant}

Considering the matrix with elements the edge weights $w_{ij}$ of the graph,
we will be particularly interested in its determinant. One can see that the conformal transformation \eqref{general_conf_transform_of_weights_Jacobson} can be written in matrix form as
\begin{equation}
W' = U W U ,
\end{equation}
with $W$ and $W'$ the corresponding matrix forms of the weights and $U$ the diagonal matrix with elements the exponentials of the conformal factors, i.e.,
\be
U_{ii} := e^{u(v_i)} .
\ee
This, in turn, implies that the determinant's transformation relation is
\begin{equation}
\det{W'} = e^{2 u(v_1)} \dots e^{2 u(v_N)} ~ \det{W} ,
\end{equation}
which may be written equivalently as
\begin{equation}
\det{W'} = J_1^2 J_2^2 \dots J_N^2  ~ \det{W} ,
\end{equation}
which makes our interest in the determinant of the weight matrix $W$ apparent from a physicist's standpoint: the conformal correlator transforms under conformal transformations as
\begin{equation}
\Braket{\Phi_1(x_1') \Phi_2(x_2') \dots \Phi_N(x_N')} = R_N \Braket{\Phi_1(x_1) \Phi_2(x_2) \dots \Phi_3(x_N)} ,
\end{equation}
where
\begin{equation}
R_N = J_1^{-2} J_2^{-2} \dots J_N^{-2} ,
\end{equation}
hinting at a description of this property via the determinant of $W$. In the following, for notational brevity, we will denote these scalar correlators with $\Sigma_N$; hence, the above would be written as
\begin{equation}\label{NcorrelatorTransformation}
\Sigma_N(x') = R_N \Sigma_N(x) ,
\end{equation}
where $x$ and $x'$ denote the set of positional variables of the correlators, i.e., $(x_1,x_2, \dots ,x_N)$ and $(x'_1,x'_2, \dots ,x'_N)$, respectively.

Transformation \eqref{conformal_factor_rescaling_explicit} allows us to introduce lengths of the underlying conformal space, which are a type of weight 
function on the edges of the graphs, into the general parameters transforming our graph, i.e., the conformal factors. This is due to the fact that we can 
express the Jacobian determinants through the graph's initial and transformed lengths by solving Eq. \eqref{conf_transform_of_lengths} for the various $j_i$. 
The system, and thus its solution, is clearly basis dependent and we have to make a choice. It is at this point that we will use the squid basis, as this choice allows for a unique solution. For example, for $N=3$ we have
\begin{equation}
\left\{
\begin{matrix}
\frac{x'_{23}}{x_{23}} = j_2 j_3  \\
\frac{x'_{12}}{x_{12}} = j_1 j_2  \\
\frac{x'_{13}}{x_{13}} = j_1 j_3
\end{matrix}
\right\}
\Rightarrow
\left\{
\begin{matrix}
\frac{x'_{23}}{x_{23}} = j_2 j_3  \\
\frac{x'_{12}}{x_{12}} \frac{x'_{13}}{x_{13}} = j_1^2 j_2 j_3
\end{matrix}
\right\}
\Rightarrow
j_1^2 = \frac{x'_{12}}{x_{12}} \frac{x'_{13}}{x_{13}} \frac{x_{23}}{x'_{23}} ,
\end{equation}
and symmetrically for the other indices. This generalizes in a straightforward manner for any $N \geq 3$, giving
\begin{align}
\begin{split}
j_1^{-2} &= \frac{x'_{23}}{x'_{12} x'_{13}} \frac{x_{12} x_{13}}{x_{23}} , \\
j_2^{-2} &= \frac{x'_{13}}{x'_{12} x'_{23}} \frac{x_{12} x_{23}}{x_{13}} , \\
j_3^{-2} &= \frac{x'_{12}}{x'_{13} x'_{23}} \frac{x_{13} x_{23}}{x_{12}} , \\
j_i^{-2} &= \frac{x'_{12} x'_{13}}{x'_{23} x_{1i}^{'2}} \frac{x_{23} x_{1i}^{2}}{x_{12} x_{13}} ~~~ \forall ~ i = 4 , \dots , N .
\label{Jacobians_func_of_lengths}
\end{split}
\end{align}
Owing to these Jacobian determinants encoding the one dilatation and $D$ special conformal transformation parameters 
we are able\footnote{A configuration where the squid graph's edges are of unit length is always available in the general case where we may apply transformations 
as denoted in \cite{bib:S1}, specifically of the form \eqref{conf_transform_of_weights_Jacobson}. This is apparent: three conformal factors can be solved for to make 
the $(123)$ triangle equilateral, and then all other conformal factors can be solved for in order to make the remaining corresponding $x_{1i}$ lengths to be of unit length. The above holds under the condition that no constraints arise due to the system's dimensionality.} to transform the $N$ lengths of the squid graph to all be unity, hence transforming 
the squid graph (more specifically the squid subgraph of the complete graph formed by all the points) to the unit squid configuration. Any other conformal transformation 
from this unit squid configuration to a new one is encoded in the values of the relevant $j_i$ by means of only the target lengths. Note that this transformation 
goes from $\tilde{x}$ to $x$ [with ${\tilde x}_{ij}\in Sq(3,N-3)$ equal to 1], something we need to reflect in the following computation \eqref{Radial_part_func_of_lengths} when applying \eqref{Jacobians_func_of_lengths}. 
The transformation coefficient of the determinant (i.e. $R_N$) can now be found to be
\begin{align}
\begin{split}
R_N &= J_1^{-2} J_2^{-2} \dots J_N^{-2} \\
&= j_1^{-2 \Delta_1} j_2^{-2 \Delta_2} \dots j_N^{-2 \Delta_N} \\
&= \left[ \frac{x_{23}}{x_{12} x_{13}} \frac{\tilde{x}_{12} \tilde{x}_{13}}{\tilde{x}_{23}} \right]^{\Delta_1} \left[ \frac{x_{13}}{x_{12} x_{23}} \frac{\tilde{x}_{12} \tilde{x}_{23}}{\tilde{x}_{13}} \right]^{\Delta_2} \left[ \frac{x_{12}}{x_{13} x_{23}} \frac{\tilde{x}_{13} \tilde{x}_{23}}{\tilde{x}_{12}} \right]^{\Delta_3} \left[ \frac{x_{12} x_{13}}{x_{23} x_{14}^{2}} \frac{\tilde{x}_{23} \tilde{x}_{14}^{2}}{\tilde{x}_{12} \tilde{x}_{13}} \right]^{\Delta_4} \dots \\
& \hspace{90mm}  \dots  \left[ \frac{x_{12} x_{13}}{x_{23} x_{1N}^{2}} \frac{\tilde{x}_{23} \tilde{x}_{1N}^{2}}{\tilde{x}_{12} \tilde{x}_{13}} \right]^{\Delta_N} \\
&= \left[ \frac{x_{23}}{x_{12} x_{13}} \right]^{\Delta_1} \left[ \frac{x_{13}}{x_{12} x_{23}} \right]^{\Delta_2} \left[ \frac{x_{12}}{x_{13} x_{23}} \right]^{\Delta_3} \left[ \frac{x_{12} x_{13}}{x_{23} x_{14}^{2}} \right]^{\Delta_4} \dots  \left[ \frac{x_{12} x_{13}}{x_{23} x_{1N}^{2}} \right]^{\Delta_N} \\
&= {x_{23}}^{-\Delta_{23}} {x_{12}}^{-\Delta_{12}} {x_{13}}^{-\Delta_{13}} {x_{14}}^{-\Delta_{14}} ~\dots~  {x_{1N}}^{-\Delta_{1N}}
\label{Radial_part_func_of_lengths}
\end{split}
\end{align}
where we define the quantities
\begin{align}
\begin{split}
\Delta_{23} &= -\Delta_1+\Delta_2+\Delta_3 +\sum_{i=4}^{N} \Delta_i , \\
\Delta_{12} &= +\Delta_1+\Delta_2-\Delta_3 -\sum_{i=4}^{N} \Delta_i , \\
\Delta_{13} &= +\Delta_1-\Delta_2+\Delta_3 -\sum_{i=4}^{N} \Delta_i , \\
\Delta_{1i} &= +2\Delta_i ~~~ \forall ~ i=4, \dots , N ,
\end{split}
\end{align}
and this form of $R_N$ is in agreement with the corresponding part of the correlator of $N$ scalar operators \cite{bib:S4}.

\section{The conformal metric}

Returning to a general transformation rule between two weight configurations \eqref{conf_transform_of_weights_Jacobson}, we can make the following postulate:
Define the \textit{metric of a conformal graph} by the transformation rule,
\begin{equation}
w_{ij} = J_i J_j \tilde{w}_{ij}\, , \hskip .5cm i\ne j
\label{conf_metric}
\end{equation}
and $w_{ii}=0$, where the quantities $J_i$ that appear in the metric are the Jacobians of the transformation to the weights $w_{ij}$ from the weights $\tilde{w}_{ij}$ of the unit squid configuration. 
As a consequence of how this metric was defined, we are guaranteed correct expressions for both the transformation of the weights and the $R_N$ coefficient.
This definition is inspired by the Weyl rescaling of the Euclidean metric in the continuum, with the unit squid being analogous to the flat metric and the metric $w_{ij}$ in the conformal class
of this unit squid being analogous to a curved, conformally flat metric.
Note that the above definition immediately implies that the invariants in \eqref{complete_set_of_generalised_conformal_invariants} can be generalized away from the geometric limit by $x_{ij}\to w_{ij}$.
Furthermore, from the definition of the metric one can easily see that
\begin{equation}
\left[ \det{W} \right]^{-1} = R_N \left[ \det{\tilde{W}} \right]^{-1}\, ,
\end{equation}
which makes it clear that the quantity $\left[ \det{W} \right]^{-1}$ is proportional to the correlator as it transforms conformally in the same way [see \eqref{NcorrelatorTransformation}],
\begin{equation}
\Sigma_N = c_N \left[ \det{W} \right]^{-1} = c_N R_N \left[ \det{\tilde{W}} \right]^{-1} = c_N R_N \Omega_N\, .
\label{relation_of_conf_correl_and_weight_det}
\end{equation}
The numerical constant $c_N$ is important in physical applications, but it will not concern us here.
The notation
\be
\Omega_N\equiv \left[ \det{\tilde{W}} \right]^{-1}
\label{ang_part_correlator_via_metric}
\ee
refers to what we will call the \textit{angular part} of the correlator, while $R_N$
will be referred to as its \textit{radial part}.

\subsection{Unit squid in general weight space}

Since $R_N$ has taken care of the covariance of the metric determinant, $\Omega_N$ has no choice but to be conformally invariant,
so it must be a function of the ${\tilde u}_{ij}$ invariants constructed from the ${\tilde w}_{ij}$ through similar expressions to \eqref{complete_set_of_generalised_conformal_invariants}. To study it, we need to write its form explicitly without sacrificing any generality. Let us start with a tentative guess and refine our approach as needed.

We may always define in a similar fashion as before, i.e., by removing any freedom granted by the conformal transformation's parameters $J_i$, the unit squid reference configuration $\tilde{w}_{ij}$. For instance, the $N=4$ unit squid metric is of the form
\begin{align}
\begin{split}
\tilde{W} &= \left[
\begin{matrix}
0 & \tilde{w}_{12} & \tilde{w}_{13} &  \tilde{w}_{14} \\
\tilde{w}_{12} & 0 & \tilde{w}_{23} & \tilde{w}_{24} \\
\tilde{w}_{13} & \tilde{w}_{23} & 0 & \tilde{w}_{34} \\
\tilde{w}_{14} & \tilde{w}_{24} & \tilde{w}_{34} & 0
\end{matrix}
\right]
=
\left[
\begin{matrix}
0 & 1 & 1 &  1 \\
1 & 0 & 1 & \tilde{w}_{24} \\
1 & 1 & 0 & \tilde{w}_{34} \\
1 & \tilde{w}_{24} & \tilde{w}_{34} & 0
\end{matrix}
\right] ,
\end{split}
\end{align}
and we immediately see that
\bea
\ut_{24} &=& \frac{\tilde{w}_{23} \tilde{w}_{14}}{\tilde{w}_{13} \tilde{w}_{24}} = \frac{1}{\tilde{w}_{24}} \Rightarrow \tilde{w}_{24} = \frac{1}{\ut_{24}} , \nonumber\\
\ut_{34} &=& \frac{\tilde{w}_{23} \tilde{w}_{14}}{\tilde{w}_{12} \tilde{w}_{34}} = \frac{1}{\tilde{w}_{34}} \Rightarrow \tilde{w}_{34} = \frac{1}{\ut_{34}}
\eea
(and similarly $\tilde{w}_{ij} = \ut_{ij}^{-1}$ for the other invariants for $N>4$), thus implying that
\begin{align}
\begin{split}
\tilde{W} =
\left[
\begin{matrix}
0 & 1 & 1 & 1 \\
1 & 0 & 1 & \ut_{24}^{-1} \\
1 & 1 & 0 & \ut_{34}^{-1} \\
1 & \ut_{24}^{-1} & \ut_{34}^{-1} & 0
\end{matrix}
\right]\, .
\label{unit_squid_metric_4p}
\end{split}
\end{align}
Note that in conventional CFT the angular part is in general an arbitrary function of the invariants \eqref{complete_set_of_generalised_conformal_invariants} which are constructed from the lengths of the edges of the complete graph; see the left sketch of Fig. \ref{fig:graphs}. Here, it is the determinant of the metric, as given in \eqref{ang_part_correlator_via_metric}, that plays the role of this function and is dependent on the set of invariant elements $\tilde{w}_{ij}$ or, equivalently, $\ut_{ij}$. These, in turn, may depend on the conventional invariants \eqref{complete_set_of_generalised_conformal_invariants} in general in an arbitrary fashion, making the two descriptions equivalent. In both descriptions the specific form that this function takes for a given correlator within a given physical system can be fixed by operations that go beyond the basic rules. In CFT it is the OPE that unlocks the fixing process. In this work, we only make some preliminary remarks regarding the fixing process in the following.

This generalizes straightforwardly to the case of the unit squid configuration of $N$ points,
\begin{align}
\begin{split}
\tilde{W} =
\left[
\begin{matrix}
0 & 1 & 1 & 1 & 1 & \dots & 1 \\
1 & 0 & 1 & \ut_{24}^{-1} & \ut_{25}^{-1} & \dots & \ut_{2N}^{-1} \\
1 & 1 & 0 & \ut_{34}^{-1} & \ut_{35}^{-1} & \dots & \ut_{3N}^{-1} \\
1 & \ut_{24}^{-1} & \ut_{34}^{-1} & 0 & \ut_{45}^{-1} & \dots & \ut_{4N}^{-1} \\
1 & \ut_{25}^{-1} & \ut_{35}^{-1} & \ut_{45}^{-1} & 0 & \dots & \ut_{5N}^{-1} \\
\vdots & \vdots & \vdots & \vdots & \vdots & \ddots & \vdots \\
1 & \ut_{2N}^{-1} & \ut_{3N}^{-1} & \ut_{4N}^{-1} & \ut_{5N}^{-1} & \dots & 0
\end{matrix}
\right]\, .
\label{unit_squid_metric_Np}
\end{split}
\end{align}
All elements are expressed in terms of the $\ut_{ij}$; therefore, they are all invariant. 

We then would like to start at this configuration to express any other weight $w_{ij}$ (in the same conformal class). However, plugging this in within the previous description 
of a reference configuration would overconstrain our description: the reference configuration in the previous is already defined to be the one whose underlying geometrical length 
space is that of the unit squid. And we cannot simply choose this over the other reference configuration, as we have already used the fact that $\tilde{x}_{ij}=1$ 
on the squid's edges when calculating \eqref{Radial_part_func_of_lengths}.

To see how we cannot constrain the system any further like this without losing generality, consider Eqs. \eqref{conf_transform_of_lengths} and \eqref{conf_transform_of_weights_Jacobson} 
and definition \eqref{bigJdef}: as we know the lengths of the edges of the initial and reference configuration, these initial lengths encode the conformal transformation 
needed toward the reference configuration. In other words, the parameters $j_i$ are known; hence, due to the system always being of some known $\Delta_i$, the parameters 
of the weight's transformation $J_i$ are also fully determined. In conclusion, once we choose that we want to move to the unit squid configuration of geometric lengths, we have also immediately and completely defined the transformation at the general weight's level. Given a general set of $w_{ij}$ (as well as $x_{ij}$, $\Delta_i$, and the convention 
that $\tilde{x}_{ij}$ is the unit squid) means that the $\tilde{w}_{ij}$ is determined and can therefore not be set by hand.

One can circumvent this problem in a number of ways, depending on what one wishes to achieve. We could set the weights $\tilde{w}_{ij}$ of the geometric unit squid $\tilde{x}_{ij}$ by hand, 
accepting the loss of generality, if we say that we wish to study this specific system of weights due to its having some desired property. In our case, however, we will keep the 
description as general as possible through the following trick: in order to preserve the result 
in \eqref{Radial_part_func_of_lengths}, i.e., without the extra factors  containing $\tilde{x}_{ij}$ appearing, 
we may pose a less strict constraint than requiring $\tilde{x}_{ij}=1$ for the squid's edges. 
Then the radial part is $R_N{\tilde R}_N^{-1}$ with
\begin{equation}
{\tilde R}_N^{-1}\equiv \left[  \frac{\tilde{x}_{12} \tilde{x}_{13}}{\tilde{x}_{23}} \right]^{\Delta_1} \left[  \frac{\tilde{x}_{12} \tilde{x}_{23}}{\tilde{x}_{13}} \right]^{\Delta_2} 
\left[  \frac{\tilde{x}_{13} \tilde{x}_{23}}{\tilde{x}_{12}} \right]^{\Delta_3} \left[  \frac{\tilde{x}_{23} \tilde{x}_{14}^{2}}{\tilde{x}_{12} \tilde{x}_{13}} \right]^{\Delta_4} \dots 
\left[  \frac{\tilde{x}_{23} \tilde{x}_{1N}^{2}}{\tilde{x}_{12} \tilde{x}_{13}} \right]^{\Delta_N}\, 
\label{length_constraint_from_radial_part}
\end{equation}
and we would like to demand that
\be
{\tilde R}_N^{-1}={\rm const}.
\ee
Now, the reference configuration's lengths need to conform to just this one constraint instead of the previous $N$. This allows for the following compromise in our 
description: we first use all the conformal transformations to achieve the form of the reference configuration's weights $\tilde{w}_{ij}$ that we desire (unit squid or 
crossing symmetric, to name two). After this, we perform a dilatation of the weights; hence,
\be
J_1= \dots =J_N=\rho ,
\ee
with $\rho$ some real positive number. 
This implies that all elements of $\tilde{W}$ are multiplied by $\rho^2$, which is invariant as a characteristic property of our reference configuration, resulting in 
its determinant being scaled by $\rho^{2N}$. At the same time, the length space is conformally transformed by the set of parameters
\be
j_i = J_i^{1/\Delta_{i}} = \rho^{-1/\Delta_{i}} ~~~ \forall ~ i \in \{1, \dots ,N\}
\ee
(note: not a dilatation), which has the overall effect of the radial part $R_N$ acquiring an 
extra factor of $\rho^{-2N}$.
In total,
\be
\Sigma_N \longrightarrow R_N {\tilde R}_N^{-1} \rho^{2N} \rho^{-2N} \Omega_N ,
\ee
and we can choose a $\rho$ such that 
\be
{\tilde R}_N^{-1} \rho^{2N}=1 ,
\ee
so as to satisfy \eqref{Radial_part_func_of_lengths} up to an irrelevant constant
that can be absorbed into $c_N$. 
To summarize, weights ${\tilde w}_{ij}$ may be applied in their 
most general form to the reference configuration while keeping the radial part of the correlator as before, as long as a general scaling factor is taken into 
account and no other assumptions are made for the lengths of the reference configuration. By doing this, any dependence of the angular part of the correlator on conformal invariants (usually in literature the invariants of the underlying length space $u,v, \dots$) is still present, only moved to all (in the general case) the intermediate quantities of the weights, i.e., ${\tilde w}_{ij}(u,v, \dots)$, from which we construct the angular part, making its dependency from the metric manifest. In other words, there is no loss of generality in \eqref{unit_squid_metric_Np},
which will be important in the following discussion.

There is one more step that has to be taken, which is to ensure that ${\rm det}\, W$ satisfies crossing symmetry, that is, to remain invariant under
the exchange of the labels $i\leftrightarrow j$. As it is, the expression for $\Sigma_N$ is not symmetric under such exchanges.
The process to make it symmetric is described in the Appendix.

\subsection{Volume considerations}

 In the following we will be concerned with the angular part $\Omega_N$ of the unit squid as it is given in \eqref{unit_squid_metric_Np}.
 Computing the determinant for $N=4$ explicitly, we obtain
\be
\label{Heron}
\Sigma_4 =  c_4\, ~ \rho^{-8} ~ x_{23}^{-\Delta_{23}} x_{13}^{-\Delta_{13}} x_{12}^{-\Delta_{12}} x_{14}^{-\Delta_{14}} \left[ 1 + \ut_{24}^{-2} + \ut_{34}^{-2} - 2 \ut_{24}^{-1} \ut_{34}^{-1} - 2 \ut_{24}^{-1} - 2 \ut_{34}^{-1} \right]^{-1}\, .
\ee
The radial part is that of the correlator of four scalar operators in a CFT, as already pointed out, but 
with the generic function\footnote{The connection between the invariants $u$ and $v$ typically used in the literature to our squid basis' invariants is 
$u=\frac{u_{24}}{u_{34}}$ and $v=u_{24}$.} $f(u,v)$ that appears in a standard conformal correlator fixed to a specific form.
As the unit squid is the representative of a conformal class, its angular part characterizes the class.
In fact, the expression in brackets in \eqref{Heron} can be recognized as Heron's formula for $-16A^2$, with $A$ the area of a triangle with edges
\be
a=\ut_{24}^{-1/2} , \quad b=\ut_{34}^{-1/2} , \quad c=1 .
\ee
A connection between the $N=4$ correlator and the volume of a tetrahedron for specific physical systems was made in \cite{Petkou}. It would be interesting to understand the form this result takes and its consequences for our approach.
If we return to the discussion we presented for $N=4$, this can actually be generalized to any $N$, which we do next. 

Looking at \eqref{unit_squid_metric_Np} we note the similarity with the Cayley-Menger determinant formula, 
allowing for an interpretation of it as an ($n$-dimensional simplex) volume. 
We remind the reader that, given $n+1$ points, the $n$-volume of the $n$-simplex (a convex polytope) defined by these 
points can be calculated via the Cayley-Menger determinant formula,
\begin{equation}
Vol_n(conv\{x_1, \dots ,x_{n+1}\})^2 = \frac{(-1)^{n+1}}{(n!)^2 2^n} CM(x_1, \dots ,x_{n+1}) \, ,
\end{equation}
where
\be
CM(x_1, \dots ,x_{n+1}) 
= \left|
\begin{matrix}
0 & 1 & 1 & 1 & \cdots & 1 \\
1 & 0 & x^2_{12} & x^2_{13} & \cdots & x^2_{1(n+1)} \\
1 & x^2_{12} & 0 & x^2_{23} & \cdots & x^2_{2(n+1)} \\
1 & x^2_{13} & x^2_{23} & 0 & \cdots & x^2_{3(n+1)} \\
\vdots & \vdots & \vdots & \vdots & \ddots & \vdots \\
1 & x^2_{1(n+1)} & x^2_{2(n+1)} & x^2_{3(n+1)} & \cdots & 0
\end{matrix}
\right|\, .
\ee
Defining
\be
q_{ij} := \ut_{ij}^{-1/2} ~,
\ee
we can write
\bea
\det{W} &=& R_N^{-1} ~ \det{\tilde{W}} \\
&=& R_N^{-1} ~ 
\left|
\begin{matrix}
0 & 1 & 1 & 1 & 1 & \cdots & 1 \\
1 & 0 & 1 & q_{24}^{2} & q_{25}^{2} & \cdots & q_{2N}^{2} \\
1 & 1 & 0 & q_{34}^{2} & q_{35}^{2} & \cdots & q_{3N}^{2} \\
1 & q_{24}^{2} & q_{34}^{2} & 0 & q_{45}^{2} & \cdots & q_{4N}^{2} \\
1 & q_{25}^{2} & q_{35}^{2} & q_{45}^{2} & 0 & \cdots & q_{5N}^{2} \\
\vdots & \vdots & \vdots & \vdots & \vdots & \ddots & \vdots \\
1 & q_{2N}^{2} & q_{3N}^{2} & q_{4N}^{2} & q_{5N}^{2} & \cdots & 0
\end{matrix}
\right|\nonumber\\
&=& R_N^{-1} ~ CM(q_2, q_3, \dots , q_N)\, .
\eea
By taking $n=N-2$, we obtain
\be
\det{W} = R_N^{-1} ~ \left[ (-1)^{N+1} ((N-2)!)^2 2^{N-2} \right]  Vol_{N-2}(conv\{q_2, \dots ,q_N\})^2
\ee
or, as a shorthand,
\begin{equation}
\det{W} = R_N^{-1} ~ \left[ (-1)^{N+1} ((N-2)!)^2 2^{N-2} \right] ~ \tilde{V}_{N-2}^2\, .
\end{equation}
One may interpret $\tilde{V}_{N-2}$ as the $(N-2)$-volume between the $N-1$ points $\{x_2,x_3, \dots ,x_N\}$ in some generalized length space where distances between points 
are given by $q_{ij}$. Note that the volume $\tilde{V}_{N-2}$ is the one corresponding to the points $2,3,\cdots, N$ specifically of the unit squid configuration (so $q_{23}^{2}=1$, etc.), and this is why it has a tilde. Since, however, this volume is a conformal invariant due to its being a property of the representative, it takes the same value in the entire conformal class, so the tilde could be dropped.

\begin{figure}[h]
    \centering
    \includegraphics[width=0.25\textwidth]{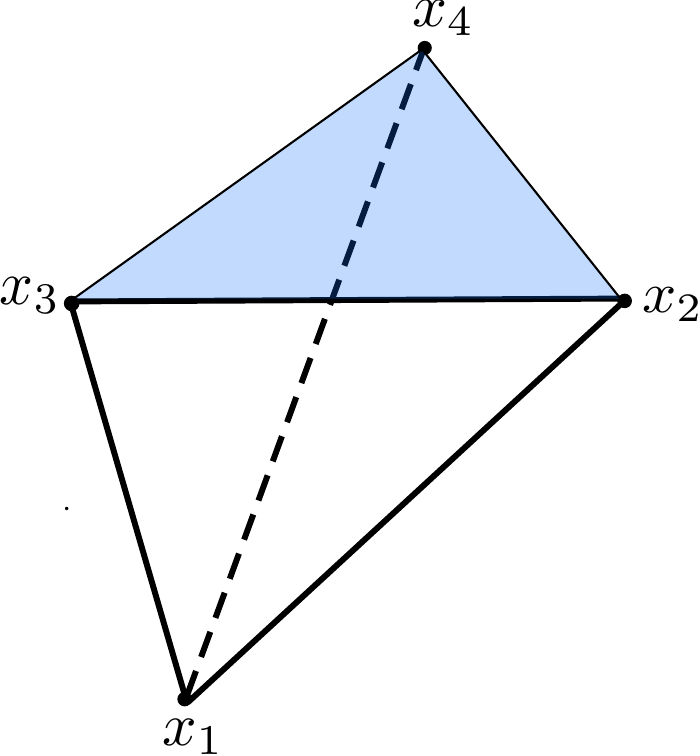}
    \caption{\label{SquidCM}The area that represents $\Omega_4$. }
\end{figure}

Putting everything together, from
\begin{equation}
\Sigma_N = c_N  \left[ \det{W} \right]^{-1}
\end{equation}
we arrive at
\begin{equation}
\Sigma_N = c_N ~ \rho^{-2N} ~ R_N ~ \frac{ (-1)^{N+1} }{ ((N-2)!)^2 2^{N-2} } ~ [ \tilde{V}_{N-2} ]^{-2} .
\end{equation}
For $N=4$ we have
\be
\Sigma_4 = - c_4 ~ \rho^{-8} ~ \frac{ 1 }{ 16 } ~ R_4 ~ \frac{ 1 }{  \tilde{V}_{2}^{2} } ,
\ee
and $\tilde{V}_{2}=A$ is the area of the tetrahedron's face that is not adjacent to $x_1$.
The interpretation of $\Omega_N$ as a volume for a general $N>3$ (area for $N=4$) is that of the discrete analog of a solid angle that the ``neck'' of the squid at $x_1$ sees, and this justifies the term angular part mentioned earlier; see Fig. \ref{SquidCM}.

\subsection{Beyond the determinant}

A possible advantage of the existence of a metric is the general statement that a metric typically contains more information than what is contained in its determinant.
One way to extract more information is to construct quantities from the metric, covariant under conformal transformations but invariant for some other reason.
These reasons may be of a physical or mathematical origin. 
The eigenvalues of the metric could be of some relevance as they transform covariantly. We do not know at this point of any physical quantity that is constructed from them
but one can have an illustrative example by demanding that they remain invariant under conformal transformations. 
This requirement yields a relation among $J_i$, $e_i$, and $\Delta_i$. 
Another example can be given based on the quantity 
$ Q= {\rm max} \{\langle f, Df\rangle : f\in C(v_i), \langle f,f \rangle=1, \langle 1,f\rangle=0\} $,
with $D$ a distance matrix and $C(v_i)$ the space of all real valued functions on the vertices \cite{bib:S2}.
On our squid it takes the value 
\be
Q = - \frac{12}{3N + \sqrt{3(N-3)(3N-8)}}
\ee
and clearly depends only on $N$. If it transforms covariantly with respect to conformal transformations, it will yield
another relation among $J_i$, $e_i$, and $\Delta_i$. 
In general, topological invariants are expected to give more such relations as long as they can be constructed from the metric.

\section{Conclusion}

Using first principles of a mathematical nature on a graph, we defined the conformal metric associated with this graph. This metric is of particular interest, both as a mathematical tool for the study of graphs under conformal transformations, which one may study simply as edge weights or in parallel to an underlying position space that transforms conformally alongside the weights, and from a physics standpoint. Regarding the latter, beyond the established immediate physical connection with the correlators of CFT through its determinant, a careful analysis of its form may provide valuable insight regarding the remaining degrees of freedom of the system that the metric encodes and their possible use. We therefore plan to explore this path in future research.

\section*{Acknowledgments}

S.K. is supported by a scholarship of the Research Committee of the National Technical University of Athens.


\setcounter{equation}{0}
\renewcommand\theequation{A\arabic{equation}}

\section*{Appendix: Crossing symmetry}

Crossing symmetry is the statement that a correlator ($\det W$) should remain invariant under a change of the vertex weight labels $\{x_i, \Delta_i\}\to \{x_j, \Delta_j\}$.
Mathematically this can be expressed by demanding that it should be invariant under the action by the $N-1$ generators of the symmetric group $S_N$.
In the squid basis it is convenient to take these as the transpositions $(1 2),\cdots ,(1 N)$. 
For example, one can generalize a noncrossing symmetry invariant (take $N=4$) as
\be\label{symmetricu}
\ut_{24}\longrightarrow \ut_{24} + \ut_{34} + \frac{1}{\ut_{24}} + \frac{1}{\ut_{34}} + \frac{\ut_{24}}{\ut_{34}} + \frac{\ut_{34}}{\ut_{24}} ,
\ee
which is simultaneously invariant under $(12)$, $(13)$, and $(14)$; therefore, it is invariant under $S_4$.

In general, the symmetrization operation of $\det W$ must be done in such a way that the product $R_N \Omega_N$ remains invariant.
One obvious strategy is to make the radial and angular parts separately invariant, the latter by substitutions as in Eq. \eqref{symmetricu}
and the former by imposing invariance on $R_N$ using the generators. Of course, this is not unique and one can move factors 
built out of the $\ut_{ij}$ between the radial and angular parts so that in the final form invariance is achieved from the cancellation of
their respective noninvariant parts.
The separate symmetrization of $R_N$ and $\Omega_N$ is not unique either. For instance, the method that uses Eq. \eqref{symmetricu} makes each element
of ${\tilde W}$ invariant, which is straightforward to carry out but is not the most general case. 
One can have a situation where the elements ${\tilde w}_{ij}$ are not invariant 
but the noninvariant parts cancel in the determinant. 
The most general discussion can be carried out in the context of the complete graph, in the left sketch of Fig. \ref{fig:graphs}, for a general $N$.
In this context, we will show how to construct a symmetric ${\rm det}\, W$.
We will illustrate some of the above statements through the $N=4$ example by showing the construction of a 
symmetric $R_4$ and then that of a symmetric $\Omega_4$ with noninvariant ${\tilde w}_{ij}$ elements. 
Then we can single out the squid part in $R_4$ and move the rest into $\Omega_4$.

For the radial part $R_N$, we note that the expressions \eqref{Jacobians_func_of_lengths} are not crossing symmetric; hence, the radial 
part \eqref{Radial_part_func_of_lengths} we presented also is not. However, by multiplying by conformal invariants formed by the lengths $x_{ij}$ in suitable powers, one can arrive at the expressions
\be
J_1 = \left[ \frac{x_{12}^2 x_{13}^2 x_{14}^2}{x_{23} x_{24} x_{34}} \right]^{\frac{\Delta_1}{6}} ~,~
J_2 = \left[ \frac{x_{12}^2 x_{23}^2 x_{24}^2}{x_{13} x_{14} x_{34}} \right]^{\frac{\Delta_2}{6}} ~,~
J_3 = \left[ \frac{x_{13}^2 x_{23}^2 x_{34}^2}{x_{12} x_{14} x_{24}} \right]^{\frac{\Delta_3}{6}} ~,~
J_4 = \left[ \frac{x_{14}^2 x_{24}^2 x_{34}^2}{x_{12} x_{13} x_{23}} \right]^{\frac{\Delta_4}{6}} ~,~
\ee
and thus at
\begin{align}
\begin{split}
R_4 =&  J_1^{-2} J_2^{-2} J_3^{-2} J_4^{-2}  \\
=& \left[ \frac{x_{23} x_{24} x_{34}}{x_{12}^2 x_{13}^2 x_{14}^2} \right]^{\frac{\Delta_1}{3}}
\left[ \frac{x_{13} x_{14} x_{34}}{x_{12}^2 x_{23}^2 x_{24}^2} \right]^{\frac{\Delta_2}{3}} 
\left[ \frac{x_{12} x_{14} x_{24}}{x_{13}^2 x_{23}^2 x_{34}^2} \right]^{\frac{\Delta_3}{3}} 
\left[ \frac{x_{12} x_{13} x_{23}}{x_{14}^2 x_{24}^2 x_{34}^2} \right]^{\frac{\Delta_4}{3}} 
\label{symm_radial_part_for_4p}\, ,
\end{split}
\end{align}
which is manifestly cross symmetric (and similarly, but with negated exponents, for the $\rho$ factor). This is generalized in a straightforward way to $N$ points. One could solve this, not for the squid basis as we did but for some basis of all lengths for which we demand crossing symmetry. Instead, and more simply, one can multiply by the set of independent conformal invariants and demand that their exponents bring the lengths under each scaling dimension of the radial part into a crossing symmetric form, i.e., have the same exponent for the length of all edges that end on the vertex in question and another exponent for all other lengths. One can then show that this form can be written as
\begin{equation}
J_i = \left[ \prod_{i \neq j \neq k \neq i} \frac{x_{ij} x_{ik}}{x_{jk}} \right]^{\frac{\Delta_i}{(N-1)(N-2)}} ,
\end{equation}
i.e., we multiply over all distinct triplets of indices which contain the index in question and for which all indices are different from each other.

To find a crossing symmetric form for the angular part $\Omega_N$, we will turn again to statement \eqref{Jacobson_prop_of_unique_rep} regarding finding 
a unique representative for a certain conformal class. 
We will prove the following statement:

The representative of \eqref{Jacobson_prop_of_unique_rep} is conformally invariant and crossing symmetric.

Applying the described property to the weights of the complete graph's representative, we find in the general case where we consider $N$ points that
\begin{align}
\begin{split}
\bar{w}_{12} \bar{w}_{13} \bar{w}_{14} \bar{w}_{15} \dots \bar{w}_{1N} &= 1 , \\
\bar{w}_{12} \bar{w}_{23} \bar{w}_{24} \bar{w}_{25} \dots \bar{w}_{2N} &= 1 , \\
\bar{w}_{13} \bar{w}_{23} \bar{w}_{34} \bar{w}_{35} \dots \bar{w}_{3N} &= 1 , \\
\bar{w}_{14} \bar{w}_{24} \bar{w}_{34} \bar{w}_{45} \dots \bar{w}_{4N} &= 1 , \\
\dots & \\
\bar{w}_{1N} \bar{w}_{2N} \bar{w}_{3N} \bar{w}_{4N} \dots \bar{w}_{(N-1)N} &= 1 ,
\end{split}
\end{align}
and, using again the definitions of the conformal invariants,
\begin{align}
\begin{split}
\bar{w}_{12} \bar{w}_{13} \bar{w}_{14} \bar{w}_{15} \dots \bar{w}_{1N} &= 1 , \\
\bar{w}_{12} \bar{w}_{23} \frac{\bar{w}_{23} \bar{w}_{14}}{\bar{w}_{13} \bar{u}_{24}} \frac{\bar{w}_{23} \bar{w}_{15}}{\bar{w}_{13} \bar{u}_{25}} \dots \frac{\bar{w}_{23} \bar{w}_{1N}}{\bar{w}_{13} \bar{u}_{2N}} &= 1 , \\
\bar{w}_{13} \bar{w}_{23} \frac{\bar{w}_{23} \bar{w}_{14}}{\bar{w}_{12} \bar{u}_{34}} \frac{\bar{w}_{23} \bar{w}_{15}}{\bar{w}_{12} \bar{u}_{35}} \dots \frac{\bar{w}_{23} \bar{w}_{1N}}{\bar{w}_{12} \bar{u}_{3N}} &= 1 , \\
\bar{w}_{14} \frac{\bar{w}_{23} \bar{w}_{14}}{\bar{w}_{13} \bar{u}_{24}} \frac{\bar{w}_{23} \bar{w}_{14}}{\bar{w}_{12} \bar{u}_{34}} \frac{\bar{w}_{23} \bar{w}_{14} \bar{w}_{15}}{\bar{u}_{45} \bar{w}_{12} \bar{w}_{13}} \dots \frac{\bar{w}_{23} \bar{w}_{14} \bar{w}_{1N}}{\bar{u}_{4N} \bar{w}_{12} \bar{w}_{13}} &= 1 , \\
\dots & \\
\bar{w}_{1N} \frac{\bar{w}_{23} \bar{w}_{1N}}{\bar{w}_{13} \bar{u}_{2N}} \frac{\bar{w}_{23} \bar{w}_{1N}}{\bar{w}_{12} \bar{u}_{3N}} \frac{\bar{w}_{23} \bar{w}_{14} \bar{w}_{1N}}{\bar{u}_{4N} \bar{w}_{12} \bar{w}_{13}} \dots \frac{\bar{w}_{23} \bar{w}_{1(N-1)} \bar{w}_{1N}}{\bar{u}_{(N-1)N} \bar{w}_{12} \bar{w}_{13}} &= 1 .
\end{split}
\end{align}
By defining $\lambda_{ij} \equiv ln(\bar{w}_{ij})$ and acting with the logarithm on the above system, we get a regular linear system with the following solution:
\begin{equation}
\begin{bmatrix}
\lambda_{23} \\
\lambda_{12} \\
\lambda_{13} \\
\lambda_{14} \\
\lambda_{15} \\
\dots \\
\lambda_{1N}
\end{bmatrix}
=
C
T
\begin{bmatrix}
0 \\
ln( \bar{u}_{24} \bar{u}_{25} \dots \bar{u}_{2N}) \\
ln( \bar{u}_{34} \bar{u}_{35} \dots \bar{u}_{3N}) \\
ln( \bar{u}_{24} \bar{u}_{34} \bar{u}_{45} \dots \bar{u}_{4N}) \\
ln( \bar{u}_{25} \bar{u}_{35} \bar{u}_{45} \dots \bar{u}_{5N}) \\
\dots \\
 ln( \bar{u}_{2N} \bar{u}_{3N} \bar{u}_{4N} \dots \bar{u}_{(N-1)N})
\end{bmatrix} ,
\label{linear_sol_ref_weights_func_of_inv}
\end{equation}
where here $C \equiv \frac{1}{(N-1)(N-2)}$ and
\be
T \equiv
\begin{bmatrix}
1 & N-2 & N-2 & -1 & -1 & \dots & -1 \\
1 & N-2 & -1 & -1 & -1 & \dots & -1 \\
1 & -1 & N-2 & -1 & -1 & \dots & -1 \\
1 & -1 & -1 & N-2 & -1 & \dots & -1 \\
1 & -1 & -1 & -1 & N-2 & \dots & -1 \\
\vdots & \vdots & \vdots & \vdots & \vdots & \ddots & \vdots \\
1 & -1 & -1 & -1 & -1 & \dots & N-2
\end{bmatrix}
~.
\ee
Knowing the weights on the squid is enough to find all the weights of the complete graph through the invariants.
This proves that the representative is conformally invariant.
More specifically, the result produced predicts, for $N=4$,
\be
\bar{W} =
\left[
\begin{matrix}
0 & \bar{u}_{24}^{1/6} \bar{u}_{34}^{-1/3} & \bar{u}_{24}^{-1/3} \bar{u}_{34}^{1/6} &  \bar{u}_{24}^{1/6} \bar{u}_{34}^{1/6} \\
\bar{u}_{24}^{1/6} \bar{u}_{34}^{-1/3} & 0 & \bar{u}_{24}^{1/6} \bar{u}_{34}^{1/6} & \bar{u}_{24}^{-1/3} \bar{u}_{34}^{1/6} \\
\bar{u}_{24}^{-1/3} \bar{u}_{34}^{1/6} & \bar{u}_{24}^{1/6} \bar{u}_{34}^{1/6} & 0 & \bar{u}_{24}^{1/6} \bar{u}_{34}^{-1/3} \\
\bar{u}_{24}^{1/6} \bar{u}_{34}^{1/6} & \bar{u}_{24}^{-1/3} \bar{u}_{34}^{1/6} & \bar{u}_{24}^{1/6} \bar{u}_{34}^{-1/3} & 0
\end{matrix}
\right] 
\equiv 
\left[
\begin{matrix}
0 & b & c &  a \\
b & 0 & a & c \\
c & a & 0 & b \\
a & c & b & 0
\end{matrix}
\right]\, .
\ee
Under any action by the generators of $S_4$, i.e., $(12),(13)$, and $(14)$,
one of the three invariant quantities $a$, $b$, or $c$ stays the same and the other two swap places, thus leaving the determinant crossing-symmetry invariant. 
Furthermore, in terms of the $a$, $b$, and $c$ parameters the determinant is
\begin{align}
\begin{split}
\label{Heron2}
\det{\bar{W}} &= a^4 + b^4 + c^4 - 2 a^2 b^2   - 2 a^2  c^2 - 2 b^2  c^2 \\
&= (a-b-c)(a+b-c)(a-b+c)(a+b+c) ,
\end{split}
\end{align}
which can be recognized as being proportional to Heron's formula for the area squared $A^2$ of a triangle with edges $a$, $b$, and $c$, giving
\be
\det{\bar{W}} = -16 A^2\, .
\ee
We also need to show that our solution is crossing symmetric in the general case where $N \geq 4$. We therefore check the metric and ultimately its 
determinant under the action of the symmetric group's generators, i.e., $(12), (13), (14), \dots , (1N)$ or, for brevity, $g_{12}, g_{13}, g_{14}, \dots , g_{1N} $, respectively. 
First, however, we will need to lay some groundwork for this. Using the definitions for the invariants we see that, under the action of the generators of the group $S_N$, the conformal invariants transform as shown in Table \ref{conformal_invariants_under_crossing_symmetry}.
\begin{table}[b]
\caption{\label{conformal_invariants_under_crossing_symmetry}
The conformal invariants when transformed under the action of the generators of the group $S_N$. The generators are listed at the top of each column (with $e$ being the identity operator). Note that $i,j,k \geq 4$ and are different from one another, as when the same index is used this signifies that the index is acted on by the group action.}
\begin{center}
\begin{tabular}{ccccc}
\hline
\hline
$e$ & $g_{12}$ & $g_{13}$ & $g_{1i}$ & $g_{1k}$  \\
\hline
$\bar{u}_{2i} $ & $ {1}/{\bar{u}_{2i}} $ & $ {\bar{u}_{2i}}/{\bar{u}_{3i}} $ & $ \bar{u}_{3i} $ & $ ({\bar{u}_{2i} \bar{u}_{3k}})/{\bar{u}_{ik}} $  \\
$\bar{u}_{3i} $ & $ {\bar{u}_{3i}}/{\bar{u}_{2i}} $ & $ {1}/{\bar{u}_{3i}} $ & $ \bar{u}_{2i} $ & $ ({\bar{u}_{3i} \bar{u}_{2k}})/{\bar{u}_{ik}} $ \\
$\bar{u}_{ij} $ & $ {\bar{u}_{ij}}/({\bar{u}_{2i} \bar{u}_{2j}}) $ & $ {\bar{u}_{ij}}/({\bar{u}_{3i} \bar{u}_{3j}}) $ & $ ({\bar{u}_{2i} \bar{u}_{3i}})/{\bar{u}_{ij}} $ & $ ({\bar{u}_{ij} \bar{u}_{2k} \bar{u}_{3k}})/({\bar{u}_{ik} \bar{u}_{jk}}) $  \\
\hline
\hline
\end{tabular}
\end{center}
\end{table}

We now move our attention to the actual weights of the metric, starting by looking at the squid's weights that are given in \eqref{linear_sol_ref_weights_func_of_inv} or, in a more explicit form, may be written as
\begin{align}
\begin{split}
\bar{w}_{23} &= a_2^{\frac{1}{N-1}} a_3^{\frac{1}{N-1}} \left[ a_4 a_5 \dots a_N  \right]^{\frac{-1}{(N-1)(N-2)}} \\
\bar{w}_{12} &= a_2^{\frac{1}{N-1}} a_3^{\frac{-1}{(N-1)(N-2)}} \left[ a_4 a_5 \dots a_N  \right]^{\frac{-1}{(N-1)(N-2)}} \\
\bar{w}_{13} &= a_2^{\frac{-1}{(N-1)(N-2)}} a_3^{\frac{1}{N-1}} \left[ a_4 a_5 \dots a_N  \right]^{\frac{-1}{(N-1)(N-2)}} \\
\bar{w}_{1i} &=  a_2^{\frac{-1}{(N-1)(N-2)}} a_3^{\frac{-1}{(N-1)(N-2)}} a_i^{\frac{1}{N-1}} \left[ a_4 a_5  \dots a_{i-1} a_{i+1} \dots a_N  \right]^{\frac{-1}{(N-1)(N-2)}} \quad \forall ~ i \in \{4, 5, \dots ,N\} ,
\end{split}
\end{align}
where we have defined
\begin{align}
\begin{split}
a_2 &\equiv \bar{u}_{24} \bar{u}_{25} \dots \bar{u}_{2N} \\
a_3 &\equiv \bar{u}_{34} \bar{u}_{35} \dots \bar{u}_{3N} \\
a_i &\equiv \bar{u}_{2i} \bar{u}_{3i} \bar{u}_{4i} \dots \bar{u}_{(i-1)i} \bar{u}_{(i+1)i} \dots \bar{u}_{Ni}     \quad \forall ~ i \in \{4, 5, \dots ,N\} .
\end{split}
\end{align}
So, we can now determine how these quantities $a_i$ transform under the generators of $S_N$ ($i\ge 4$),
{\allowdisplaybreaks
\begin{align}
g_{12} a_2 &= \frac{1}{a_2} , \nonumber \\
g_{12} a_3 &=  \frac{a_3}{a_2} , \nonumber \\
g_{12} a_i &= \frac{1}{\bar{u}_{2i}} \frac{\bar{u}_{3i}}{\bar{u}_{2i}} \frac{\bar{u}_{4i}}{\bar{u}_{24} \bar{u}_{2i}} \dots
\frac{\bar{u}_{(i-1)i}}{\bar{u}_{2(i-1)} \bar{u}_{2i}}  \frac{\bar{u}_{(i+1)i}}{\bar{u}_{2(i+1)} \bar{u}_{2i}}  \dots  \frac{\bar{u}_{Ni}}{\bar{u}_{2N} \bar{u}_{2i}} \nonumber \\
&=
\frac{a_i}{a_2}  \frac{1}{\bar{u}_{2i}^{(N-2)}} ,  \nonumber \\
g_{13} a_2 &= g_{13} (\bar{u}_{24} \bar{u}_{25} \dots \bar{u}_{2N}) = \frac{a_2}{a_3} , \nonumber \\
g_{13} a_3 &= g_{13} (\bar{u}_{34} \bar{u}_{35} \dots \bar{u}_{3N}) = \frac{1}{a_3} , \nonumber \\
g_{13} a_i &= \frac{\bar{u}_{2i}}{\bar{u}_{3i}} \frac{1}{\bar{u}_{3i}} \frac{\bar{u}_{i4}}{\bar{u}_{3i} \bar{u}_{34}} \dots \frac{\bar{u}_{(i-1)i}}{\bar{u}_{3i} \bar{u}_{3(i-1)}} \frac{\bar{u}_{(i+1)i}}{\bar{u}_{3i} \bar{u}_{3(i+1)}} \dots \frac{\bar{u}_{iN}}{\bar{u}_{3i} \bar{u}_{3N}}   \nonumber \\
&=
\frac{a_i}{a_3}  \frac{1}{\bar{u}_{3i}^{(N-2)}} ,  \nonumber \\
g_{1i} a_2 &= \frac{\bar{u}_{24} \bar{u}_{3i}}{\bar{u}_{4i}} \dots  \frac{\bar{u}_{2(i-1)} \bar{u}_{3i}}{\bar{u}_{(i-1)i}} \bar{u}_{3i}  \frac{\bar{u}_{2(i+1)} \bar{u}_{3i}}{\bar{u}_{(i+1)i}} \dots \frac{\bar{u}_{2N} \bar{u}_{3i}}{\bar{u}_{Ni}} \nonumber \\
&= \frac{a_2}{a_i} \bar{u}_{3i}^{(N-2)} , \nonumber \\
g_{1i} a_3 &=  \frac{\bar{u}_{34} \bar{u}_{2i}}{\bar{u}_{4i}}  \dots  \frac{\bar{u}_{3(i-1)} \bar{u}_{2i}}{\bar{u}_{(i-1)i}} \bar{u}_{2i} \frac{\bar{u}_{3(i+1)} \bar{u}_{2i}}{\bar{u}_{(i+1)i}}   \dots \frac{\bar{u}_{3N} \bar{u}_{2i}}{\bar{u}_{Ni}}  \nonumber \\
&= \frac{a_3}{a_i} \bar{u}_{2i}^{(N-2)} , \nonumber \\
g_{1i} a_i &=  \frac{(\bar{u}_{2i} \bar{u}_{3i})^{(N-2)}}{a_i} , \nonumber \\
g_{1j} a_i &=  \frac{(\bar{u}_{2j} \bar{u}_{3j})^{(N-2)}}{\bar {u}_{ij}^{(N-2)}} \frac{a_i}{a_j} .
\end{align}
}
Using these transformation rules, one can explicitly verify via the solution that
\begin{equation}
\begin{aligned}
g_{12} \bar{w}_{12} &= \bar{w}_{12} , & g_{12} \bar{w}_{13} &= \bar{w}_{23} , & g_{12} \bar{w}_{23} &= \bar{w}_{13} , \\
g_{13} \bar{w}_{12} &= \bar{w}_{23} , & g_{13} \bar{w}_{13} &= \bar{w}_{13} , & g_{13} \bar{w}_{23} &= \bar{w}_{12} , \\
g_{23} \bar{w}_{12} &= \bar{w}_{13} , & g_{23} \bar{w}_{13} &= \bar{w}_{12} , & g_{23} \bar{w}_{23} &= \bar{w}_{23} ,
\end{aligned}
\end{equation}
and using these and the transformation laws of the invariants in Table \ref{conformal_invariants_under_crossing_symmetry} it follows that
\begin{align}
\begin{split}
g_{12} \bar{w}_{1i} = \bar{w}_{2i} ,\hskip .5cm  g_{12} \bar{w}_{2i}  = \bar{w}_{1i} ,\hskip .5cm  g_{12} \bar{w}_{3i} = \bar{w}_{3i} ,\hskip .5cm   g_{12} \bar{w}_{ij} = \bar{w}_{ij} ,\\
g_{13} \bar{w}_{1i} = \bar{w}_{3i} ,\hskip .5cm  g_{13} \bar{w}_{2i}  = \bar{w}_{2i} ,\hskip .5cm g_{13} \bar{w}_{3i} = \bar{w}_{1i} ,\hskip .5cm   g_{13} \bar{w}_{ij} = \bar{w}_{ij} ,\\
g_{23} \bar{w}_{1i} = \bar{w}_{1i} ,\hskip .5cm  g_{23} \bar{w}_{2i}  = \bar{w}_{3i} ,\hskip .5cm  g_{23} \bar{w}_{3i}  = \bar{w}_{2i} ,\hskip .5cm  g_{23} \bar{w}_{ij} = \bar{w}_{ij} ,\\
g_{1i} \bar{w}_{12} = \bar{w}_{2i} ,\hskip .5cm  g_{1i} \bar{w}_{13}  = \bar{w}_{3i} ,\hskip .5cm  g_{1i} \bar{w}_{23}  = \bar{w}_{23} ,\hskip .5cm g_{1i} \bar{w}_{ij} = \bar{w}_{1j} ,\\
g_{1k}\bar{w}_{ij} = \bar{w}_{ij} .
\end{split}
\end{align}
That is, $S_N$ acts covariantly on the indices of ${\bar W}$, as it should.
At this point, having proven that the indices of the $\bar{w}_{ij}$ are exchanged in the expected way under all generators and thus any element of $S_N$, 
we are immediately also able to prove the determinant $\det{\bar{W}}$ crossing symmetric. Consider the weights in $\bar{W}$ under the action of $g_{ij}$: 
this induces an exchange between the relevant columns and also an exchange between the relevant rows. Thus, the determinant will pick up two negative 
signs from this exchange, resulting in the same determinant value in the end. And with this, the proposed angular part is proven crossing symmetric.

The previous treatment of the angular part is done in general terms, invoking the use of the $\bar{u}_{ij}$ invariants which, in turn, depend on the weights $\bar{w}_{ij}$. One may, however, wish to express all quantities through the lengths and scaling dimensions of the problem. We will therefore also outline an example of an alternative approach. In particular, we take $N=4$ and make the following ansatz for an element of the weights $m_{ij}$ of the required reference configuration:
\be
m_{14} = u_{24}^{a_1 \Delta_1 + a_2 \Delta_2 + a_3 \Delta_3 + a_4 \Delta_4 } u_{34}^{b_1 \Delta_1 + b_2 \Delta_2 + b_3 \Delta_3 + b_4 \Delta_4 } .
\ee
Here, the invariants are constructed from the given lengths of the problem. We have allowed for eight unknown coefficients in the exponents, however, by repeatedly applying the generators $g_{12}$, $g_{13}$, and $g_{14}$ as before on this, we not only get all other elements of the associated weight matrix $M$ but also restrict the unknown coefficients down to two, namely, $a_1$, and $a_2$. Ultimately, one finds that
\begin{align}
\begin{split}
\label{symm_weights_from_ansatz}
m_{12} &=  u_{24}^{ + a_1 \Delta_1 +a_1 \Delta_2+ a_2 \Delta_3 + a_2 \Delta_4 }  u_{34}^{ -2a_1 \Delta_1 -2 a_1 \Delta_2-2 a_2 \Delta_3 -2 a_2 \Delta_4 } , \\
m_{13} &=  u_{24}^{-2 a_1 \Delta_1 -2 a_2 \Delta_2 -2 a_1 \Delta_3 -2 a_2 \Delta_4 }  u_{34}^{+ a_1 \Delta_1  + a_2 \Delta_2 +a_1 \Delta_3+ a_2 \Delta_4 } , \\
m_{14} &= u_{24}^{a_1 \Delta_1 + a_2 \Delta_2 + a_2 \Delta_3 + a_1 \Delta_4 } u_{34}^{a_1 \Delta_1 + a_2 \Delta_2 + a_2 \Delta_3 + a_1 \Delta_4 } , \\
m_{23} &=  u_{24}^{ + a_2 \Delta_1 + a_1 \Delta_2+a_1 \Delta_3 + a_2 \Delta_4 }  u_{34}^{+ a_2 \Delta_1+a_1 \Delta_2  +a_1 \Delta_3  + a_2 \Delta_4} , \\
m_{24} &=  u_{24}^{ -2 a_2 \Delta_1 -2 a_1 \Delta_2-2 a_2 \Delta_3 -2 a_1 \Delta_4 } u_{34}^{ + a_2 \Delta_1 +a_1 \Delta_2+ a_2 \Delta_3 + a_1 \Delta_4 } , \\
m_{34} &=   u_{24}^{+ a_2 \Delta_1  + a_2 \Delta_2 +a_1 \Delta_3+ a_1 \Delta_4 } u_{34}^{-2 a_2 \Delta_1 -2 a_2 \Delta_2 -2 a_1 \Delta_3 -2 a_1 \Delta_4 } ,
\end{split}
\end{align}
and from this we can derive
\begin{align}
\begin{split}
\label{Heron3}
\det{M} =&
+ m_{14}^2 m_{23}^2 + m_{13}^2 m_{24}^2 + m_{12}^2 m_{34}^2 \\
&- 2 m_{14} m_{23} m_{13} m_{24}  - 2 m_{14} m_{23}  m_{12} m_{34}  -2 m_{13} m_{24} m_{12} m_{34} \\
=&
+u_{24}^{ 2 E } u_{34}^{ 2 E }
+u_{24}^{ -4 E } u_{34}^{ +2 E }
+u_{24}^{ +2 E } u_{34}^{ -4 E } 
-2 u_{24}^{ - E } u_{34}^{ +2 E }
-2  u_{24}^{ +2 E } u_{34}^{ - E }
-2  u_{24}^{ - E } u_{34}^{ - E } ,
\end{split}
\end{align}
where we have defined
\begin{equation}
E \equiv (a_1 + a_2)( \Delta_1 +  \Delta_2+ \Delta_3 +  \Delta_4) .
\end{equation}
Note that \eqref{Heron3} is once again proportional to Heron's formula, as in \eqref{Heron} and \eqref{Heron2}, for a triangle with edges
\be
u_{24}^{ E/2 } u_{34}^{ E/2 } , \quad u_{24}^{ E/2 } u_{34}^{ -E } , \quad u_{24}^{ -E } u_{34}^{ E/2 } .
\ee
As a last note on the angular part, if we require \eqref{symm_weights_from_ansatz} to conform to the property given by \eqref{Jacobson_prop_of_unique_rep}, this would imply taking $a_1=a_2$.


Having shown both the radial and angular parts to be able to be written in a cross symmetric way, we may wish to nevertheless work in some other 
basis of lengths in the radial part, like the squid we have defined. We can do this while simultaneously preserving crossing symmetry simply by choosing 
the lengths we wish to keep in the radial part, which should be the covariant part of the correlator, and move what remains, which therefore should be some 
invariant expression, into the angular part by multiplying and dividing by this invariant quantity. As an example, let us split out of $R_4$ [see \eqref{symm_radial_part_for_4p}], the radial part of the squid $R_{4,sq}$, as in \eqref{Radial_part_func_of_lengths}. This produces the expression
\begin{equation}
R_4 = R_{4,sq} u_{24}^{-\frac{1}{3} \Delta_1 +\frac{2}{3} \Delta_2 -\frac{1}{3} \Delta_3 +\frac{2}{3} \Delta_4 } u_{34}^{-\frac{1}{3} \Delta_1 -\frac{1}{3} \Delta_2 +\frac{2}{3} \Delta_3 +\frac{2}{3} \Delta_4 } ,
\end{equation}
and the remaining invariant part can be redistributed to the angular part $\Omega_N$.

\newpage

\end{document}